# Static de-Sitter black holes abhor charged scalar hair

Yu-Ping An[1,2,a] , Li Li[1,2,3,4,b]

[1] CAS Key Laboratory of Theoretical Physics, Institute of Theoretical Physics, Chinese Academy of Sciences, Beijing 100190, China
[2] School of Physical Sciences, University of Chinese Academy of Sciences, Beijing 100049, China
[3] School of Fundamental Physics and Mathematical Sciences, Hangzhou Institute for Advanced Study, UCAS, HangZhou 310024, China
[4] Peng Huanwu Collaborative Center for Research and Education, Beihang University, Beijing 100191, China

**Abstract** We prove a no charged scalar hair theorem for static black holes in de-Sitter spacetime in the region between the event horizon and the cosmological horizon. The proof does not depend on the assumption of spherical symmetry. It allows for general non-minimal coupling functions of the scalar field to gravity and electromagnetic fields, and for higher curvature term corrections to Einstein gravity. The extension to other asympitotic spacetimes is applicable by requiring appropriate boundary conditions. Our result excludes the possibility for spontaneous scalarization of charged scalar around static charged de-Sitter black holes.

## 1 Introduction

It is well-known that all asymptotically flat stationary electro-vacuum black holes in four spacetime dimensions are characterized by their mass, angular momentum and electric charge—the Kerr–Newman metric (see [1] for a review). This celebrated uniqueness theorem leads to the conjecture that the final state of gravitational collapse in the presence of any type of matter and energy is solely characterized by mass, angular momentum and electric charge, without any other parameters known as "hair". This is called the no hair conjecture. Following this conjecture, various no hair theorems were established [2–5]. Thus, the (stationary) black holes were understood to be remarkably simple objects. Nevertheless, all existing no hair theorems need certain assumptions, such as spherical or axial symmetry, appropriate boundary conditions or specific coupling and potential form of matter fields. As time passes by, various counterexamples with different kinds of hair have been found, including the black holes with skyrmion [6,7], Yang-Mills [8–11], axion [12–15], and dilaton hair [16,17]. Indeed, various hairy black holes can develop if one considers new couplings. For example, scalarized black holes can form an infinite (countable) number of branches with large freedom in their multipole structure by considering an electromagnetic non-minimal coupling in asymptotically flat spacetime [18].

At a more fundamental level, by taking quantum effects into consideration, a black hole should have enormous "quantum hair" inspired particularly by the Bekenstein–Hawking entropy and the black hole information paradox. There are some new couplings arising by considering quantum corrections, such as a Gauss–Bonnet coupling. Moreover, many alternative theories beyond general relativity have appeared to improve our understanding in various aspects of gravity, such as spacetime singularity, dark energy, dark matter and inflation. Therefore, the physics of black holes has been far fuller and richer than previously believed. It might be more reasonable to consider no hair theorems as constrains for possible stationary classical field configurations. For a recent review of no hair theorems, see [19].

Among all no hair theorems, there is a strong one concerning the static charged scalar field proved by Mayo and Bekenstein [20]. The authors claimed that *there exists no non-extreme static and spherical charged black hole endowed with hair in the form of a charged scalar, whether minimally or non-minimally coupled to gravity, and with a regular positive semi-definite self-interaction potential*, assuming additionally asymptotic flat boundary. Nevertheless, the authors of [21] pointed out a flaw in Bekenstein's argument and presented a counterexample recently. Besides, according to the astronomical observation [22,23], our universe could be equipped with a positive cosmological constant. The positive cosmological constant changes the spacetime from asymptotic flat to asymptotic de-Sitter. In particular, there should be a cosmological horizon that significantly changes

---

[a] e-mail: anyuping@itp.ac.cn (corresponding author)
[b] e-mail: liliphy@itp.ac.cn

Springer

the boundary conditions, and therefore could invalidate the Bekenstein's argument. Thus far, it was only shown that the static and spherical de-Sitter black can not take charged scalar hair in the Abelian–Higgs model [24]. Therefore, it is interesting and important to understand the no hair theorems in more general cases.

In this paper, we will prove that *there exists no static and charged de-Sitter black hole endowed with hair in the form of a charged scalar field*. We shall consider black holes with finite temperature. We use the method developed in [24,25] and assumes no other symmetries except that the spacetime is static between the event horizon and the cosmological horizon. Therefore, our results are valid for both isotropic and anisotropic cases. Besides, our results are also valid for spacetimes with non-minimally coupled scalar fields and higher curvature terms. The extension to other asymptotic spacetimes is possible if one imposes appropriate boundary conditions for matter fields. After carefully inspecting the boundary conditions for asymptotic flat case, our results are found to be consistent with [21] where a charged black hole with non-trivial scalar hair was constructed in asymptotically flat case. Moreover, there have been some discussions about the spontaneous scalarization of charged black holes [26,27] and of de-Sitter black holes [28]. Our result would exclude the existence of spontaneous scalarization of charged scalar around a large class of static de-Sitter black holes.

## 2 The theory and setup

We consider a complex charged scalar field $\Psi$ non-minimally coupled to gravity, together with a Maxwell field $A_\mu$:

$$S = \frac{1}{2\kappa_N^2} \int d^4x \sqrt{-g} [\mathcal{P}(|\Psi|^2)\mathcal{L}_G + \mathcal{L}_M],$$
$$\mathcal{L}_M = -(\mathcal{D}_\mu \Psi)^* \mathcal{D}^\mu \Psi - V(|\Psi|^2) - \frac{Z(|\Psi|^2)}{4} F^{\mu\nu} F_{\mu\nu}, \tag{2.1}$$

where $F_{\mu\nu} = \nabla_\mu A_\nu - \nabla_\nu A_\mu$ and $\mathcal{D}_\mu = \nabla_\mu - iqA_\mu$ with $q$ the charge of the scalar field. $\mathcal{P}$, $Z$ and $V$ are arbitrary smooth functions of $|\Psi|^2$, but we require $Z \geq 0$ to ensure positivity of the kinetic term for $A_\mu$. $\mathcal{L}_G$ can be any scalar invariant of differential transformation composed by metric tensor, which means our results would apply to gravity theory with any number of arbitrary higher derivative terms, such as the Gauss–Bonnet term.

The equations of motion (EOMs) for matter fields are given by

$$\mathcal{D}_\mu \mathcal{D}^\mu \Psi = (\dot{V}(|\Psi|^2) + \frac{\dot{Z}(|\Psi|^2)}{4} F^{\mu\nu} F_{\mu\nu} - \dot{\mathcal{P}}(|\Psi|^2)\mathcal{L}_G)\Psi, \tag{2.2}$$

$$\nabla^\mu [Z(|\Psi|^2) F_{\mu\nu}] = iq(\Psi^* \mathcal{D}_\nu \Psi - \Psi(\mathcal{D}_\nu \Psi)^*), \tag{2.3}$$

where dots represent derivatives with respect to $|\Psi|^2$. Einstein's equation is omitted since we will not use it in this work.

To proceed further, we impose some assumptions on the geometry. Here we consider static spacetime with a Killing vector field $\xi^\mu$ which is timelike between the event horizon and the cosmological horizon (if there is one). It norm $\xi^\mu \xi_\mu = -\beta^2$ vanishes at both horizons which are Killing horizon associated with $\xi^\mu$. Moreover, $\xi^\mu$ is hypersurface orthogonal and therefore satisfies the Frobenius condition

$$\xi_{[a} \nabla_b \xi_{c]} = 0. \tag{2.4}$$

We define, for later convenience, the projector to the space-like hypersurface $\Sigma$ that is orthogonal to $\xi^\mu$.

$$h^\mu_\nu = \delta^\mu_\nu + \beta^{-2} \xi^\mu \xi_\nu, \tag{2.5}$$

which projects vectors to $\Sigma$. We denote $D_\mu$ as the induced connection (covariant derivative) on $\Sigma$.

Since we are interested in the static case, for scalar field and Maxwell field, we require that their Lie derivatives with respect to $\xi^\mu$ vanish, i.e.

$$\mathcal{L}_\xi \Psi = 0, \quad \mathcal{L}_\xi A_\mu = 0. \tag{2.6}$$

The condition (2.6) seems natural at first sight. However, since we are considering a theory with $U(1)$ gauge symmetry, the Lagrangian (2.1) is invariant under the local gauge transformation

$$\Psi \to \Psi e^{i\alpha}, \quad A_\mu \to A_\mu + \frac{1}{q} \nabla_\mu \alpha, \tag{2.7}$$

with $\alpha$ an arbitrary function. Therefore, we need to consider the gauge degree of freedom carefully. For this purpose, we separate the norm and the phase factor of the scalar field and define a gauge invariant vector field

$$\Psi = \psi e^{i\alpha_0}, \quad B_\mu = A_\mu - \frac{1}{q} \nabla_\mu \alpha_0. \tag{2.8}$$

One might wonder what if this gauge is singular, which could happen when there are topological defects such as vortex lines in spacetime. We postpone the discussion of this issue until Sect. 5, but the spoiler is that this does not affect our final conclusion.

Now we impose the constrains on the gauge invariant quantities $\psi$ and $B^\mu$:

$$\mathcal{L}_\xi \psi = 0, \quad \mathcal{L}_\xi B_\mu = 0. \tag{2.9}$$

Let $t$ to be the time coordinate adapted to the Killing vector $\xi^\mu$, such that $\xi = \frac{\partial}{\partial t}$. The above condition implies that $\psi$ and each component of $B^\mu$ are independent of time $t$. One can choose $\alpha$ arbitrarily as long as $A_\mu$ is changed accordingly. Especially, if one chooses $\alpha_0 = \omega_0 t$ with $\omega_0$ a constant, $A_\mu = B_\mu + \omega_0/q (dt)_\mu$ is also independent of $t$, which is the case considered in [21]. Though $\Psi = \psi e^{i\omega_0 t}$ now depends on $t$ explicitly, which seems to violate the original assumption (2.6), the later one (2.9) still holds.

We are ready to show a no-go theorem for the existence of static and charged de-Sitter black hole endowed with hair in the form of a charged scalar.

## 3 Spherical symmetric case

We begin with the simple case with spherical symmetry and minimal coupling between scalar and gravity, i.e. $\dot{\mathcal{P}} = 0$. Thanks to the symmetry of the system, the static charged black hole with charged scalar hair reads

$$ds^2 = -f(r)e^{-\chi(r)}dt^2 + \frac{dr^2}{f(r)} + r^2 d\Omega^2,$$
$$\Psi = \psi(r), \quad B = B_t(r)dt + B_r(r)dr, \tag{3.10}$$

where $B_\mu$ is the gauge invariant vector field defined in (2.8) and all functions only depend on the radial coordinate $r$. Besides the event horizon $H^+$, there is cosmological horizon outside $H^+$ due to the presence of the positive cosmological constant. We denote the location of the event horizon and cosmological horizon as $r_H$ and $r_c$, respectively. In particular, the blackening function $f$ is vanishing at both horizons and is positive between them.

However, as one can easily verify, the $r$-component of Maxwell's equations (2.3) yields $B_r = 0$ for non-trivial $\psi(r)$. Therefore, the only non-vanishing components of $B_\mu$ is $B_t(r)$. Then the EOMs (2.2) and (2.3) become

$$f\psi'' + \frac{q^2}{f}e^\chi B_t^2 \psi = (\dot{V} + \frac{\dot{Z}}{2}e^\chi B_t'^2)\psi, \tag{3.11a}$$

$$(ZB_t')' = \frac{q^2}{f}B_t\psi^2, \tag{3.11b}$$

where prime stands for derivative with respect to $r$. Since the event horizon is an inner horizon relative to the cosmological horizon, one wonders if the no-inner horizon theorem [29, 30] could apply to the present case. Unfortunately, similar argument fails because the blackening function $f$ is positive between $r_H$ and $r_c$.

Note that the blackening function $f = 0$ on both horizons. To have a smooth horizon, it is clear from the above equations of motion that either $\psi$ or $B_t$ should vanish on the horizon. Moreover, if $\psi = 0$ on the horizon, it must vanish everywhere near the horizon (see Appendix B in [29] for more details). We have only required $\psi$ to be analytic in a neighborhood of the horizon. Therefore, one obtains $B_t(r_H) = B_t(r_c)$ for the black hole with non-trivial charged scalar hair $\psi$. Multiplying Eq. (3.11b) by $B_t$ and integrating by part, we get

$$ZB_t B_t'|_{r_H}^{r_c} = \int_{r_H}^{r_c} dr \left( \frac{q^2}{f} B_t^2 \psi^2 + ZB_t'^2 \right). \tag{3.12}$$

We see that the left hand side is vanishing while the right hand side is non-negative. Therefore, when the scalar hair is non-vanishing, one can only require $B_t = 0$ between the event horizon and the cosmological horizon, and the hairy black hole dose not take electric charge. On the other hand, if the black hole does not support scalar hair $\psi$, the resulting solution could be the Reissner-Nördstrom-de Sitter (RNdS) black hole. Hence, the static de-Sitter black hole cannot take both the electric and charged scalar hair, which is the no-hair result.

We have shown from Eq. (3.12) that $B_t = 0$ to have a non-trivial scalar hair, for which Eq. (3.11a) becomes

$$f\psi'' = \dot{V}\psi. \tag{3.13}$$

We multiply it by $\psi$ and integrate it by part. Then we obtain

$$f\psi\psi''|_{r_H}^{r_c} = \int_{r_H}^{r_c} dr(f\psi'^2 + \dot{V}\psi^2). \tag{3.14}$$

Note that the left hand side is zero since $f(r_c) = f(r_H) = 0$. Therefore, if $\dot{V} \geq 0$ we have $\psi = 0$. Moreover, this argument applies to a real scalar $\phi$ with a potential $V_0(\phi)$. One can find that there is no scalar hair once

$$\phi \frac{dV_0(\phi)}{d\phi} \equiv \phi V_0'(\phi) \geq 0. \tag{3.15}$$

It was proved that [24] static spherical black holes in de-Sitter cannot support the real scalar $\phi$ in convex potentials, i.e. $V_0''(\phi) \geq 0$. Our result (3.15) is different from $V_0''(\phi) \geq 0$. They have overlap but are not equivalent to each other. Note that we did not need to use the Einstein's equation.

In summary, for the static spherical charged black hole, we find that there is no charged scalar hair in the region between the event and the cosmological horizons. Moreover, for the neutral black hole case, it dose not support complex scalar $\psi$ (neutral scalar $\phi$) if its potential $\dot{V} \geq 0$ ($\phi V_0'(\phi) \geq 0$).

## 4 Generic case without spherical symmetry

We now generalize the no scalar hair theorem to the case without spherical symmetry. In order to analyse the problem from a geometric point of view and to reduce reliance on metric ansatz as much as possible, we take a similar procedure as in [24] where the EOMs are projected to the hypersurface $\Sigma$. Before getting into details of the EOMs, we quote a formula used in [24].

$$D_\alpha(\beta\omega^{\alpha\mu\nu\cdots}) = \beta(\nabla_\alpha\Omega^{\alpha\mu'\nu'\cdots})h^\mu_{\mu'}h^\nu_{\nu'}\ldots, \quad (4.16)$$

for an antisymmetric tensor $\Omega$ whose Lie derivative with respect to $\xi^\mu$ vanishes, where $\omega$ is the $\Sigma$-projection of $\Omega$. Proof of this formula is presented in A.

We first define some notations for later convenience.

$$\Phi = \frac{\xi_\mu}{\beta}B^\mu, \quad a_\mu = h^\nu_\mu B_\nu,$$
$$e^\mu = \frac{\xi_\nu}{\beta}F^{\mu\nu}, \quad f_{\mu\nu} = h^\sigma_\mu h^\rho_\nu F_{\sigma\rho}. \quad (4.17)$$

Since $\Sigma$ is spacelike, the norms of $a_\mu$, $e^\mu$ and $f_{\mu\nu}$ are non-negative. Let's consider Maxwell equation (2.3) first. Projected by $h^\nu_\mu$ and multiplied by $\beta$, it becomes

$$\beta h^\mu_\alpha \nabla_\nu(Z(\psi^2)F^{\nu\alpha}) = 2\beta q^2\psi^2 a^\mu. \quad (4.18)$$

From Eq. (4.16), one can easily find that

$$D_\mu(\beta Z(\psi^2)f^{\mu\nu}) = \beta h^\nu_\alpha \nabla_\mu(Z(\psi^2)F^{\mu\alpha}). \quad (4.19)$$

Then Eq. (4.18) becomes

$$D_\mu(\beta Z(\psi^2)f^{\mu\nu}) = 2\beta q^2\psi^2 a^\nu. \quad (4.20)$$

Contracting Eq. (4.20) with $a_\nu$ and integrating by part, we get

$$\int_{\partial\Sigma}\beta Z(\psi^2)f^{\mu\nu}a_\nu n_\mu$$
$$= \int_\Sigma \left(\frac{1}{2}\beta Z(\psi^2)f^{\mu\nu}f_{\mu\nu} + 2\beta q^2\psi^2 a^\nu a_\nu\right), \quad (4.21)$$

where $n_\mu$ is the normal vector of $\partial\Sigma$. The right hand side of above equation is over a sum of squares. Since $Z(\psi^2)$, $B_\mu$ and $F^{\mu\nu}$ appear in the energy momentum tensor

$$T_{\mu\nu} = \partial_\mu\psi\partial_\nu\psi + q^2 B_\mu B_\nu \psi^2 + \frac{Z(\psi^2)}{2}F_{\mu\alpha}F_\nu{}^\alpha + \frac{1}{2}g_{\mu\nu}\mathcal{L}_M, \quad (4.22)$$

they should have bounded norm at both horizons, and the same for their spatial components $f_{\mu\nu}$ and $a_\mu$ (see more arguments in [24]). By Schwarz inequality,

$$|f^{\mu\nu}a_\nu n_\mu|^2 \leq (a^\mu a_\mu)(f^{\alpha\nu}n_\alpha f^\beta_\nu n_\beta). \quad (4.23)$$

The right hand side of the inequality is obviously finite, therefore the second line of Eq. (4.21) is finite. Since $\beta = 0$ on a horizon, both sides of Eq. (4.21) are vanishing. This means $f_{\mu\nu} = 0$ and $a^\nu = 0$ (if $q, \psi \neq 0$) between two horizons. Note that this result does not depend on any spatial symmetries. All we need is the static condition.

There is still a component of Maxwell equation we have not analysed. We contract the Maxwell equation with $\xi_\nu/\beta$ and get

$$\frac{\xi_\nu}{\beta}\nabla_\mu(Z(\psi^2)F^{\mu\nu}) = 2q^2\psi^2\Phi. \quad (4.24)$$

One can verify $\frac{\xi_\nu}{\beta}\nabla_\mu(Z(\psi^2)F^{\mu\nu}) = D_\mu(Z(\psi^2)e^\mu)$ using Eq. (2.4). So we have

$$D_\mu(Z(\psi^2)e^\mu) = 2q^2\psi^2\Phi. \quad (4.25)$$

We multiply the above equation by $\beta\Phi$ and again integrate by part. We then obtain

$$\int_{\partial\Sigma}\beta Z(\psi^2)\Phi e^\mu n_\mu$$
$$= \int_\Sigma(2q^2\beta\psi^2\Phi^2 + Z(\psi^2)e^\mu D_\mu(\beta\Phi)),$$
$$= \int_\Sigma(2q^2\beta\psi^2\Phi^2 + Z(\psi^2)e^\mu(\beta e_\mu + h^\nu_\mu\mathcal{L}_\xi B_\nu)),$$
$$= \int_\Sigma(2q^2\beta\psi^2\Phi^2 + Z(\psi^2)\beta e^\mu e_\mu). \quad (4.26)$$

For similar reason as above, the surface integral is zero for asymptotic de-Sitter case, while the last line is positive semi-definite. This gives $e^\mu = 0$ and $\Phi = 0$ (if $q, \psi \neq 0$). Again, this requires only the static condition and the no-hair result follows.

Thus far, from the Maxwell equation (2.3), we have shown that for a static asymptotic de-Sitter spacetime $F_{\mu\nu} = 0$ and $B_\mu = 0$ if the charged scalar $\Psi$ is non-vanishing. To put it another way, this is equivalent to that no charged scalar hair can exist as long as the gauge vector field is non-trivial, which is the main content of our no scalar hair theorem. We should point out that the coupling between $\Psi$ and $A_\mu$ via $\mathcal{D}_\mu\Psi = (\nabla_\mu - iqA_\mu)\Psi$ is crucial in our proof. If the charge $q = 0$, there would be no gauge invariant combination $B_\mu$ (2.8), and one can always choose a gauge for $A_\mu$ so that the boundary conditions stated above are potentially violated. A typical example for this case is the Reissner-Nördstrom-de Sitter black hole.

Now we turn to the scalar equation (2.2). For the static case, we have shown that $F_{\mu\nu} = 0$ and $B_\mu = 0$ to support a non-vanishing $\psi$. Hence, the equation becomes

$$\nabla_\mu\nabla^\mu\psi - \left[\dot{V}(\psi^2) - \dot{\mathcal{P}}(\psi^2)\mathcal{L}_G\right]\psi = 0. \quad (4.27)$$

From Eq. (2.9) we have $\mathcal{L}_\xi \psi = \xi^\nu \nabla_\nu \psi = 0$ and obtain

$$\mathcal{L}_\xi \nabla_\mu \psi = \xi^\nu \nabla_\nu \nabla_\mu \psi + \nabla_\nu \psi \nabla_\mu \xi^\nu,$$
$$= \nabla_\mu (\xi^\nu \nabla_\nu \psi) = 0. \quad (4.28)$$

We multiply both sides of Eq. (4.27) and use Eq. (4.16) to get

$$D_\mu(\beta D^\mu \psi) = \beta(\dot{V}(\psi^2) - \dot{\mathcal{P}}(\psi^2)\mathcal{L}_G)\psi. \quad (4.29)$$

Multiplying this equation by $\psi$ and integrating over the spacelike region $\Sigma$ between the two horizons, we have

$$\int_{\partial\Sigma} \beta\psi D^\mu\psi n_\mu = \int_\Sigma (\beta D_\mu\psi D^\mu\psi + \beta(\dot{V}(\psi^2) - \dot{\mathcal{P}}(\psi^2)\mathcal{L}_G)\psi^2). \quad (4.30)$$

The boundary condition for the surface integral is similar to that of the Maxwell equation. Generally $\mathcal{L}_G$ is not positive definite or negative definite, so we cannot say much if the scalar field has a non-minimal coupling with gravity. A static hairy black hole with a conformally coupled scalar field can be found in [31]. Nevertheless, if the coupling is minimal, i.e. $\dot{\mathcal{P}}(\psi^2) = 0$, then we only need to require $\dot{V}(\psi^2) \geq 0$ which guarantees that the second line of Eq. (4.30) is positive semi-definite. Therefore, there exists no scalar hair when $\dot{V}(\psi^2) \geq 0$ and the scalar field is minimally coupled to gravity.

Another way to construct positive definite terms was done in [24]. Multiplying both sides of Eq. (4.29) by $\dot{V}(\psi^2)\psi$ and integrating, we have

$$\int_{\partial\Sigma} \beta\dot{V}(\psi^2)\psi D^\mu\psi n_\mu$$
$$= \int_\Sigma (\beta(2\ddot{V}(\psi^2)\psi^2 + \dot{V}(\psi^2))D_\mu\psi D^\mu\psi$$
$$+ \beta(\dot{V}(\psi^2) - \dot{\mathcal{P}}(\psi^2)\mathcal{L}_G)\dot{V}(\psi^2)\psi^2). \quad (4.31)$$

If we denote $V_0(\psi) = V(\psi^2)$, then we have $\dot{V}(\psi^2)\psi = V_0'(\psi)$ and $2\ddot{V}(\psi^2)\psi^2 + \dot{V}(\psi^2) = V_0''(\psi)$, where prime stands for the derivative with respect to $\psi$. Then the above equation becomes

$$\int_{\partial\Sigma} \beta V_0'(\psi) D^\mu\psi n_\mu$$
$$= \int_\Sigma (\beta V_0''(\psi) D_\mu\psi D^\mu\psi + \beta V_0'(\psi)^2), \quad (4.32)$$

for minimal coupling with gravity ($\dot{\mathcal{P}} = 0$). For the potential convex, i.e. $V_0''(\psi) \geq 0$, one arrives at the the no-hair result.

One might be aware that the discussion about the scalar equation is equivalently to the neutral scalar (i.e. $q = 0$) as considered in [24,25]. Moreover, it was shown that the no neutral-scalar-hair result also holds for stationary axisymmetric case [25]. Note that $\dot{V}(\psi^2) \geq 0$ is equivalent to $\psi V_0'(\psi) \geq 0$. The convex potential (i.e. $V_0''(\psi) \geq 0$) and the potential satisfying $\psi V_0'(\psi) \geq 0$ have non-empty intersection but do not include each other. This is the second part of our no scalar hair theorem for static de-Sitter spacetime. Again, we do not require any spatial symmetry in the proof.

For the neutral scalar case, one cannot gauge away the phase factor anymore. So the scalar field may have a time dependent phase $\Psi \sim e^{i\omega t}$, which violates the condition (2.6), but still keeps the energy-momentum tensor independent of $t$ (see e.g. [32–35]).

## 5 Conclusion and discussion

We have provided a no charged scalar hair theorem for static de-Sitter black holes under the static condition (2.8). Thanks to the presence of the cosmological horizon, we have shown that there exists no charged scalar hair in the region between the event and cosmological horizons of a static charged de-Sitter black hole, no matter the scalar field is minimally coupled to gravity or not. Moreover, if the black hole has vanishing vector hair, we have shown that it cannot support scalar fields when the scalar potential satisfies $V_0''(\psi) \geq 0$ or $\psi V_0'(\psi) \geq 0$ for the scalar minimally coupled to gravity.

These results can be extended to other asymptotic spacetimes provided the following conditions are satisfied.

$$\int_{\partial\Sigma} \beta Z(\psi^2) f^{\mu\nu} a_\nu n_\mu = 0,$$
$$\int_{\partial\Sigma} \beta Z(\psi^2) \Phi e^\mu n_\mu = 0,$$
$$\int_{\partial\Sigma} \beta \psi D^\mu \psi n_\mu = 0. \quad (5.33)$$

Therefore, for other cases without the cosmological horizon, as long as all fields die off sufficiently fast so that (5.33) goes to zero at $\partial\Sigma$ (typically at infinity), our main conclusion still holds. For charged black holes in asymptotically flat and anti-de Sitter spacetimes, the above conditions at the boundary are not satisfied in general. So one cannot exclude charged scalar hair in asymptotic flat and anti-de Sitter charged black holes. Indeed, a spherical charged black hole in asymptotically flat case with non-trivial scalar hair was constructed in [21]. In anti-de Sitter case, the charged black hole with spontaneous scalarization of charged scalar is known as holographic superconductors (see [36] for a review).

One may worry about the issue of topological defects such as vortex lines. Indeed, when vortex lines are presented, the phase $\alpha_0$ of the scalar field $\Psi$ would jump from 0 to $2\pi$ at somewhere. Therefore, the gauge of Eq. (2.8) may present a delta function in $B_\mu$ where $\alpha_0$ jumps. This singularity does not affect our discussion of the bulk integrals since it does not change the semi-positive definiteness of these integrals. So we only need to check if the boundary conditions can be broken. The delta function appears only in $B_\mu$ and $F_{\mu\nu}$ is gauge invariant, so the second and third conditions of Eq. (5.33) are not influenced. To check the first condition, let's choose a cylinder-like coordinate system where the jump of $\alpha_0$ happens at $\theta = 0$. For convenience, we only present singular parts. Then, we have $B_\mu \sim B_\theta \sim 2\pi\delta(\theta)$, and also $a_\mu \sim a_\theta \sim 2\pi\delta(\theta)$. Plugging this singular term into the first condition of Eq. (5.33), we get

$$\int_{\partial\Sigma} \beta Z(\psi^2) f^{\mu\nu} a_\nu n_\mu \sim \int_{\partial\Sigma} dz d\theta \beta Z(\psi^2) f^{\mu\theta} 2\pi\delta(\theta) n_\mu,$$
$$\sim \int dz (2\pi\beta Z(\psi^2) f^{\mu\theta} n_\mu)|_{\theta=0}. \quad (5.34)$$

We see that the integral is still vanishing since $\beta = 0$, so the first condition is still satisfied even with topological defects.

From our analysis, one can easily see that our results also apply to non-minimal coupling between $\Psi$ and $A_\mu$ in the form $(\mathcal{D}_\mu\Psi)^*\mathcal{D}^\mu\Psi \to \nabla^\mu\psi\nabla_\mu\psi - P(\psi^2)B^2$ with the Stückelberg function $P(\psi^2) \geq 0$, where $\psi$ and $B_\mu$ are defined in Eq. (2.8). This kind of coupling, together with $Z(|\Psi|^2)F^2$ and $\mathcal{P}(|\Psi|^2)R_{GB}^2$ are non-minimal couplings considered frequently in studies of black hole spontaneous scalarization [26–28], where $R_{GB}^2$ stands for the Gauss–Bonnet term. Hence our results suggest a no-go theorem for the spontaneous scalarization of charged scalar field in charged static de-Sitter black holes. In order to obtain spontaneous scalarization in charged static de-Sitter black holes, one should consider a neutral scalar with a reasonable choice of scalar potential.

More recently, it has been shown that the number of horizons of static black holes can be constrained by energy conditions of matter fields [37]. If the energy momentum tensor $T_{\mu\nu}$ in the interior of a static black hole satisfies strong energy condition or null energy condition, there is at most one inner horizon behind the event horizon. The argument was based on the Einstein's equation $G_{\mu\nu} = T_{\mu\nu}$ and is difficult to generalize to other gravity theories, such as higher derivative gravity and the case with non-minimal couplings between metric and matter field. Nevertheless, our theorem can be used to exclude the second inner horizon of the static charged black hole with charged scalar hair for the theory (2.1). Suppose there is a second inner horizon, then there exists a connected region inside the black hole event horizon such that the Killing vector $\xi^\mu$ is timelike between the two inner horizons and is null at both horizons. These two inner horizons play the same role of the event and cosmological horizons in the present work. Following the similar steps as in Sect. 4, we can exclude the second inner horizon so as to support non-trivial charged scalar hair.

Our theorem might have some significance for the strong cosmic censorship. The formation of a charged black hole by gravitational collapse necessitates the presence of a charged sector on top of the Einstein–Maxwell system. It has been recently argued that the strong cosmic censorship can be violated under the linear charged scalar perturbation of RNdS black holes [38–40]. Nevertheless, once the backreaction to the geometry is considered, no matter how weak the backreaction is, we have shown that the resulting black hole, if static, cannot take charged scalar hair.

We are aware that our results may not apply to matter fields with multiple components charged under a single $U(1)$ gauge field, including multiple scalar fields and vector fields. This is due to the fact that matter fields with multiple components may have relative phase between different components, which cannot be eliminated by gauge choice if we have only one $U(1)$ gauge symmetry. These relative phases would present in the second line of Eqs. (4.21) and (4.26) as non-positive definite factors, and therefore invalidate our analysis. To study these cases, one should appeal to other methods. It is also interesting to generalize our discussion to a stationary black hole.

**Acknowledgements** This work was partially supported by the National Natural Science Foundation of China Grants No.12122513, No.12075298, No.11991052 and No.12047503, and by the Chinese Academy of Sciences Project for Young Scientists in Basic Research YSBR-006.

**Data Availability Statement** This manuscript has no associated data or the data will not be deposited. [Authors' comment: No datasets were generated or analysed during the current study.]

## A Proof of Eq. (4.16)

In this appendix, we will prove Eq. (4.16) that appears in the main text. We point out that the spacetime is static for which the associated Killing vector $\xi$ is hypersurface orthogonal.

$$D_\alpha(\beta\omega^{\alpha\mu\nu\cdots}) = h^\alpha_\lambda h^\mu_{\mu'} h^\nu_{\nu'} \ldots \nabla_\alpha(\beta\Omega^{\lambda\mu'\nu'\cdots}),$$
$$= h^\alpha_\lambda h^\mu_{\mu'} h^\nu_{\nu'} \ldots (\beta\nabla_\alpha \Omega^{\lambda\mu'\nu'\cdots} + \Omega^{\lambda\mu'\nu'\cdots}\nabla_\alpha\beta).$$
(A.1)

Notice that $\beta = \sqrt{-\xi^\mu\xi_\mu}$, $h^\alpha_\lambda = \delta^\alpha_\lambda + \frac{\xi_\lambda\xi^\alpha}{\beta^2}$, and $h^\alpha_\lambda \nabla_\alpha\beta = \nabla_\lambda\beta$. The above equation becomes

$$h^\mu_{\mu'} h^\nu_{\nu'} \ldots \left(\beta\nabla_\alpha \Omega^{\alpha\mu'\nu'\cdots} + \frac{\xi^\alpha\xi_\lambda}{\beta}\nabla_\alpha \Omega^{\lambda\mu'\nu'\cdots} + \Omega^{\alpha\mu'\nu'\cdots}\nabla_\alpha\beta\right)$$
$$= h^\mu_{\mu'} h^\nu_{\nu'} \ldots \left(\beta\nabla_\alpha \Omega^{\alpha\mu'\nu'\cdots} + \frac{\xi^\alpha\xi_\lambda}{\beta}\nabla_\alpha \Omega^{\lambda\mu'\nu'\cdots} - \Omega^{\alpha\mu'\nu'\cdots}\frac{\xi_\lambda\nabla_\alpha\xi^\lambda}{\beta}\right).$$
(A.2)

The first term is what we want. So we only need to show that the last two terms are vanishing.

We note that that the Lie derivative of $\Omega$ with respect to $\xi^\mu$ is zero, i.e.

$$\mathcal{L}_\xi \Omega^{\alpha\mu\nu\cdots} = \xi^\lambda \nabla_\lambda \Omega^{\alpha\mu\nu\cdots} - \Omega^{\lambda\mu\nu\cdots}\nabla_\lambda\xi^\alpha$$
$$- \Omega^{\alpha\lambda\nu\cdots}\nabla_\lambda\xi^\mu - \Omega^{\alpha\mu\lambda\cdots}\nabla_\lambda\xi^\nu - \cdots$$
$$= 0,$$
(A.3)

from which the two terms in parenthesis of Eq. (A.2) become

$$\frac{\xi_\lambda}{\beta}(\xi^\alpha \nabla_\alpha \Omega^{\lambda\mu'\nu'\cdots} - \Omega^{\alpha\mu'\nu'\cdots}\nabla_\alpha\xi^\lambda)$$
$$= \frac{\xi_\lambda}{\beta}(\Omega^{\lambda\alpha\nu'\cdots}\nabla_\alpha\xi^{\mu'} + \Omega^{\lambda\mu'\alpha\cdots}\nabla_\alpha\xi^{\nu'} + \cdots). \quad \text{(A.4)}$$

If the first index of $\Omega$ anti-commutes with other indices, especially if $\Omega$ is antisymmetric, we will obtain

$$\Omega^{\lambda\ldots\alpha\cdots}\xi_\lambda \nabla_\alpha\xi^\mu$$
$$= \Omega^{\lambda\ldots\alpha\cdots}\xi_\lambda \left(h^{\alpha'}_\alpha - \frac{\xi_\alpha\xi^{\alpha'}}{\beta^2}\right)\nabla_{\alpha'}\xi^\mu,$$
$$= \Omega^{\lambda\ldots\alpha\cdots}\xi_\lambda D_\alpha\xi^\mu - \Omega^{\lambda\ldots\alpha\cdots}\xi_\lambda\xi_\alpha \frac{\xi^{\alpha'}\nabla_{\alpha'}\xi^\mu}{\beta^2},$$
$$= \Omega^{\lambda\ldots\alpha\cdots}\xi_\lambda D_\alpha\xi^\mu.$$
(A.5)

However, this term will be contracted with $h^{\mu'}_\mu$ in the end. Since $D_\mu$ is the induced connection of $h^{\mu'}_\mu$, they commute with each other, i.e. $D_\nu h^{\mu'}_\mu = 0$. Moreover, from the definition of $h^{\mu'}_\mu$ (2.5), we find $h^{\mu'}_\mu \xi^\mu = 0$. Therefore, we immediately have $h^{\mu'}_\mu D_\alpha \xi^\mu = 0$. Plugging Eq. (A.5) into Eq. (A.4) and contracting with $h^\mu_{\mu'} h^\nu_{\nu'} \ldots$, we get

$$h^\mu_{\mu'} h^\nu_{\nu'} \ldots \left(\frac{\xi^\alpha\xi_\lambda}{\beta}\nabla_\alpha \Omega^{\lambda\mu'\nu'\cdots} - \Omega^{\alpha\mu'\nu'\cdots}\frac{\xi_\lambda\nabla_\alpha\xi^\lambda}{\beta}\right)$$
$$= h^\mu_{\mu'} h^\nu_{\nu'} \ldots \frac{\xi_\lambda}{\beta}(\Omega^{\lambda\alpha\nu'\cdots}D_\alpha\xi^{\mu'} + \Omega^{\lambda\mu'\alpha\cdots}D_\alpha\xi^{\nu'} + \cdots),$$
$$= 0.$$
(A.6)

This relation, together with Eq. (A.2), gives our desired result in Eq. (4.16).